\documentclass[preprint,showpacs,preprintnumbers,amsmath,amssymb]{revtex4}

\usepackage{graphicx}
\usepackage{dcolumn}
\usepackage{bm}

\begin{document}

\preprint{APS/}

\title{An integrable optical-fiber source of polarization
entangled photon-pairs in the telecom band}

\author{Xiaoying Li}
\altaffiliation[Now at School of Precision Instrument and
Opto-electronics Engineering, Tianjin University, Tianjin, 300072,
P. R. China. ] {Email address: xiaoyingli@tju.edu.cn}

\author{Chuang Liang, Kim Fook Lee, Jun Chen, Paul L. Voss, and Prem Kumar}%

\affiliation{%
Center for Photonic Communication and Computing, EECS Department,
Northwestern University, 2145 Sheridan Road, Evanston, IL
60208-3118, USA
}%

\date{\today}

\begin{abstract}
We demonstrate an optical-fiber based source of polarization
entangled photon-pairs with improved quality and efficiency, which
has been integrated with off-the-shelf telecom components and is,
therefore, well suited for quantum communication applications in
the 1550\,nm telecom band. Polarization entanglement is produced
by simultaneously pumping a loop of standard dispersion-shifted
fiber with two orthogonally-polarized pump pulses, one propagating
in the clockwise and the other in the counter-clockwise direction.
We characterize this source by investigating two-photon
interference between the generated signal-idler photon-pairs under
various conditions. The experimental parameters are carefully
optimized to maximize the generated photon-pair correlation and to
minimize contamination of the entangled photon-pairs from
extraneously scattered background photons that are produced by the
pump pulses for two reasons: i) spontaneous Raman scattering
causes uncorrelated photons to be emitted in the signal/idler
bands and ii) broadening of the pump-pulse spectrum due to
self-phase modulation causes pump photons to leak into the
signal/idler bands. We obtain two-photon interference with
visibility $>90$\% without subtracting counts caused by the
background photons (only dark counts of the detectors are
subtracted), when the mean photon number in the signal (idler)
channel is about 0.02/pulse, while no interference is observed in
direct detection of either the signal or the idler photons.
\end{abstract}

\maketitle

\section{Introduction}

Entangled photon-pairs are a critical resource for realizing
various quantum information protocols, such as quantum
teleportation, quantum cryptography, database searching, clock
synchronization, and quantum computing~\cite{Bennett98}.
Therefore, the efficient generation and distribution of quantum
entanglement is of prime importance. The vast majority of
entangled-photon sources that are in use in various laboratories
around the world today rely on spontaneous parametric down
conversion in $\chi ^{(2)}$ crystals~\cite{Weinberg70,Kwait95}.
However, formidable engineering challenges remain in coupling the
entangled photons into standard optical
fibers~\cite{Bovino03,Castelletto04,Andrews04} for transmission,
storage, and manipulation over long distances. Therefore, an
efficient and compact entangled photon-pair source created from
the fiber itself and emitting entangled photon-pairs in the
low-loss 1550\,nm telecom band is desirable.

Recently, fiber based sources of entangled photon-pairs have been
developed by exploiting the $\chi ^{(3)}$ (Kerr) nonlinearity of
the
fiber~\cite{Fiorentino02,Li05a,Takesue04,Li04,Rarity05,Sharping04,Wang05}.
When the pump wavelength is close to the zero-dispersion
wavelength of the fiber, phase-matching is achieved and the
probability amplitude for inelastic four-photon scattering (FPS)
is significantly enhanced. In this process, two pump photons at
frequency $\omega _p$ scatter through the Kerr nonlinearity of the
fiber to create an energy-time entangled signal and idler
photon-pair at frequencies $\omega _s$ and $ \omega _i$,
respectively, such that $2\omega _p=\omega _s+\omega _i$. Because
of the isotropic nature of the Kerr nonlinearity in
fused-silica-glass fiber, the scattered, correlated photon-pairs
are predominantly co-polarized with the pump photons. By
coherently adding two such orthogonally-polarized FPS processes,
polarization entanglement has been created as
well~\cite{Li05a,Takesue04,Kumar05}. Using such a fiber based
source, storage and long-distance distribution of polarization
entanglement over 50\,km of standard single-mode fiber with
negligible decoherence has also been recently
presented~\cite{Li05b}, which demonstrates the viability of
all-fiber sources for use in quantum memories and quantum logic
gates. However, for the polarization entangled photon-pair source
reported in Ref.~\onlinecite{Li05a}, the relative phase between
the two relatively delayed pump pulses needed to be tracked and
locked in order to obtain a stable entangled state for the
photon-pairs, which made the system somewhat complicated. On the
other hand, such phase locking is not needed for the source
reported in Refs.~\onlinecite{Takesue04} and~\onlinecite{Kumar05},
in which a loop of dispersion-shifted fiber (DSF) connected to the
two output ports of a polarization beam splitter (PBS) is
simultaneously pumped from clockwise (CW) and counter-clockwise
(CCW) directions with two orthogonally-polarized pump pulses. In
this scheme, polarization entanglement is produced by a coherent
combination of the two FPS processes induced by the CW and CCW
propagating pump pulses. More importantly, this setup has two
additional advantages: i) the cross-polarized spontaneous Raman
scattering (RS), which is a source of background photons, is
automatically suppressed~\cite{Li04} and ii) the correlated
photon-pairs generated via FPS from both the CW and CCW pumps
emerge in the forward direction, whereas half of the photon-pairs
generated by each of the two pump pulses are lost in the backward
direction in the Sagnac-loop scheme of Ref.~\onlinecite{Li05a}.

Despite the relative pros and cons of the two
schemes~\cite{Li05a,Takesue04,Kumar05}, the fiber-based sources
demonstrated thus far have been quite inefficient for several
reasons. Firstly, the quality of the generated
entanglement~\cite{Li05a,Takesue04} has been limited by the
extraneously scattered background photons, whose origin is
two-fold: i) spontaneous Raman scattering from the pump pulses
causes uncorrelated Stokes (anti-Stokes) photons to be emitted in
the signal (idler) bands; ii) broadening of the pump-pulse
spectrum due to self-phase modulation (SPM) in the fiber causes
pump photons to leak into the signal/idler bands. Secondly, the
quantum correlation of the photon-pairs produced via FPS has not
been maximized for lack of careful consideration of the
differences in the spectral shape of the pulsed pump and that of
the filters used in the signal and idler bands~\cite{Chen05}.
Finally, the filters used to isolate the pumps [free-space
gratings, fiber Bragg gratings (FBGs), and array waveguide
gratings (AWGs)], have been quite lossy resulting in a low rate
for the generated entangled photon-pairs.

In this paper, using a scheme similar to that reported in
Refs.~\cite{Takesue04,Kumar05}, we demonstrate an optical-fiber
source of polarization entangled photon-pairs with improved
quality and efficiency, which has been integrated with
off-the-shelf telecom components and is, therefore, well suited
for quantum communication applications in the 1550\,nm telecom
band. The quality of the source is improved by carefully
optimizing the experimental parameters in order to suppress the
contamination due to RS and SPM of the pump pulses and to maximize
the correlation of the signal-idler photon-pairs produced via FPS.
The efficiency of the source is improved by using cascaded WDM
filters (CWDMF) that not only isolate the pump with lesser loss
but also improve the correlation of the generated signal-idler
photon-pairs because of their squarish (super-Gaussian) shape. We
characterize the source by experimentally investigating two-photon
interference (TPI) between the generated signal-idler photon-pairs
under various conditions. TPI with visibility $>90$\% is obtained
without subtracting counts caused by the background photons (only
dark counts of the detectors are subtracted), while no
interference is observed in direct detection of either the signal
or the idler photons. All four Bell states can be created with our
source.

\section{Experimental Details}

\begin{figure}
\includegraphics{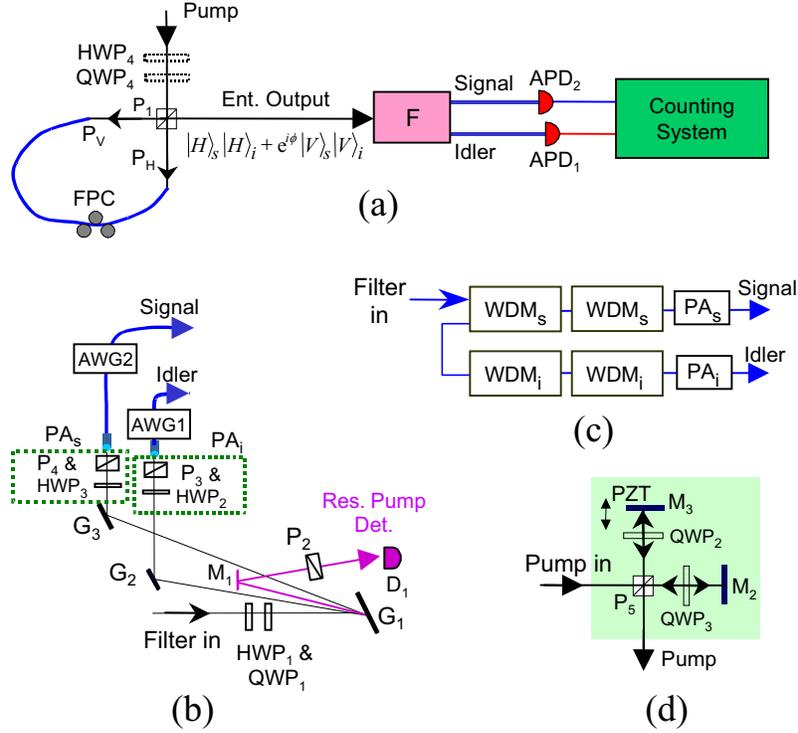}
\caption{\label{setup} (a) Experimental setup. (b) Configuration
of the filter F composed of a double-grating filter followed by an
AWG (DGFAWG). (c) Configuration of the filter F composed of two
cascaded WDM filters (CWDMF). (d) Schematic of the pumping scheme.
P$_1$--P$_5$, polarization beam splitters; F, filter;
G$_1$--G$_3$, diffraction gratings; AWG, array-waveguide grating;
M$_1$--M$_5$, mirrors; FPC, fiber polarization controller; QWP,
quarter-wave plate; HWP, half-wave plate; PA$_s$ and PA$_i$,
polarization analyzers.}
\end{figure}

A schematic of our experimental setup is shown in
Fig.~\ref{setup}(a). Signal and idler photon-pairs at wavelengths
of 1533.9\,nm and 1543.5\,nm, respectively, are produced via the
FPS process in a 300\,m piece of DSF having a zero-dispersion
wavelength at $\lambda_0=1538\pm 2$\,nm. Pump pulses with a
central wavelength of 1538.7\,nm are decomposed into horizontally
and vertically polarized components, $P_H$ and $P_V$,
respectively, by use of a polarization beam splitter (PBS) P$_1$
and launched into the DSF. $P_H$ propagates CW in the DSF
producing correlated photon-pairs $|H\rangle_s |H \rangle_i$,
whereas $P_V$ propagates CCW producing $|V \rangle_s |V
\rangle_i$. A fiber polarization controller (FPC) is used with the
DSF to compensate the birefringence introduced by bending and
coiling of the fiber on a spool. It takes about 2\,$\mu$s for a
pump pulse to propagate through the loop formed by the DSF and
P$_1$. For such a short time period, the pump pulses $P_H$ and
$P_V$, launched simultaneously into the loop in CW and CCW
directions, can be viewed as propagating along a frozen path,
irrespective of the local environmental perturbations, acoustic
and thermal, which occur on millisecond time scales. The
correlated wave packets $| H\rangle_s| H \rangle_i$ and $| V
\rangle_s | V \rangle_i$ also see a similar frozen path and are
coherently superimposed upon arrival at P$_1$. For a photon-pair
emerging at the output port of P$_1$, there is no way to determine
whether it was created by the CW (horizontally polarized) or the
CCW (vertically polarized) pulse [see Fig.~\ref{setup}(a)]. This
indistinguishability gives rise to polarization entanglement,
resulting in the output state $|\Psi \rangle =| H\rangle _s|
H\rangle _i + e^{i\phi }| V\rangle _s| V\rangle _i$, where $\phi$
is the relative phase difference between the wave packets $|
H\rangle_s| H \rangle_i$ and $| V \rangle_s | V \rangle_i$. In our
experiment $\phi=2\phi_p $, where $\phi_p $ is the relative phase
between $P_H$ and $P_V$.

This source can produce all four polarization-entangled Bell
states. When $\phi_p=0, \frac{\pi}{2}$, the states $|\Psi^\pm
\rangle = |H\rangle_s |H\rangle_i \pm |V\rangle_s |V\rangle_i$ are
created. In our setup, $\phi_p=0$ is obtained by launching a
linearly polarized pump pulse and then using a half-wave plate
(HWP$_4$) to rotate the polarization direction to $45^{\circ }$
relative to P$_1$. Additionally, when a quarter-wave plate
(QWP$_4$) with its axis parallel to P$_1$ is placed right after
HWP$_4$, then $\phi_p= \frac{\pi}{2}$ is achieved. The other two
Bell states $|\Phi^\pm\rangle = |H\rangle_s |V\rangle_i \pm
|V\rangle_s |H\rangle_i$ can be prepared by inserting a properly
oriented HWP in the idler channel. Non-maximally entangled pure
states having an arbitrary degree of polarization entanglement can
also be created with this setup by choosing the two pump pulses
($P_H$ and $P_V$) to have unequal powers, which can be
accomplished by rotating the HWP$_4$.

Polarization entanglement is measured by using polarization
analyzers (PAs) [each composed of a rotatable half-wave plate
(HWP) and a PBS, see Fig.~\ref{setup}(b)] in the signal and idler
channels. For the state $| \Psi \rangle =| H \rangle _s| H
\rangle_i +e^{i2\phi_p }| V \rangle_s | V \rangle_i $, when the
polarization analyzers in the signal and idler channels are set to
$\theta_1$ and $\theta_2$, respectively, the single-count
probability for the signal and idler photons is $R_j = \frac
12\eta _j \alpha$ $(j=1,2)$. The coincidence-count probability
$R_{12}$ can then be expressed as
  $R_{12} = \frac{1}{2} \xi \eta _1 \eta _2 \alpha [ \cos ^2\theta _1\cos
^2\theta _2+\sin ^2\theta _1\sin ^2\theta _2 \nonumber
 + 2\cos (2\phi _p)\sin \theta _1\cos \theta _1\sin \theta _2\cos
\theta _2 ]$, where $\eta _j$ ($j=1,2$) is the total detection
efficiency, $\theta _j$ ($j=1,2$) is the angle of the PA in each
channel, $\alpha$ is the photon-pair production rate, and $\xi$ is
a coefficient determined by the spectrum of the pump pulse and
that of the filters in the signal and idler bands.

\begin{figure}
\includegraphics[width=3.25in]{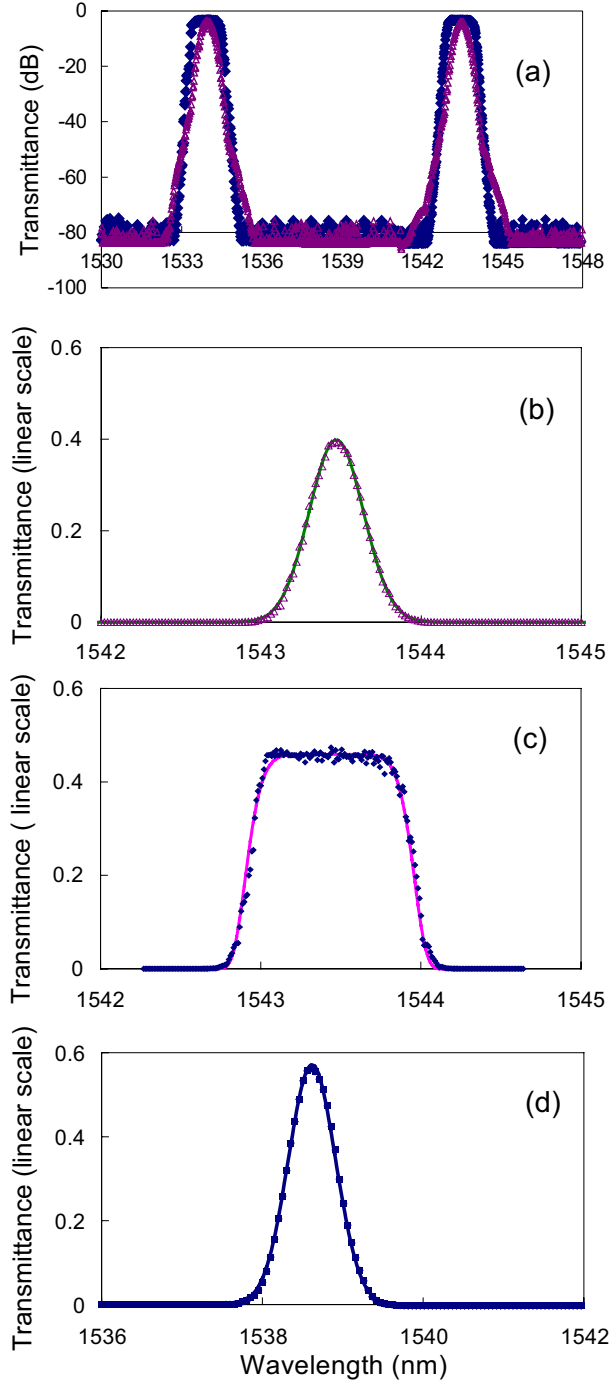}
\caption{\label{spectra} Passband spectra for the various filters
used in our setup. Solid curves are fits to the data. (a)
Transmission spectra for the DGFAWG filter (purple triangles) and
the CWDMF filter (blue diamonds). (b) One passband of the DGFAWG
fitted with a Gaussian function $f(\lambda )=0.4\exp [-\frac
{1}{2} (\frac{\lambda -1543.5}{0.17})^2]$, where the FWHM is about
0.4\,nm. (c) One passband of the CWDMF fitted with a
supper-Gaussian function $f(\lambda )=0.46\exp [-\frac
12(\frac{\lambda -1543.5}{0.49})^{2m}]$, $m=4.1$, where the FWHM
is about 1.0\,nm. (d) Filtered spectrum of the pump pulse fitted
with a Gaussian function $f(\lambda )=0.55\exp
[-\frac{1}{2}(\frac{\lambda -1538.7}{0.33})^2]$, where the FWHM is
about 0.8\,nm.}
\end{figure}

To detect the scattered photon-pairs with good signal-to-noise, an
isolation of the pump photons in excess of 100\,dB is required.
Because of the non-zero response time of the Kerr
nonlinearity~\cite{Li05c}, there are actually three kinds of
photons that emerge from the output port [see
Fig.~\ref{setup}(a)]: (i) entangled signal and idler photon-pairs
produced by the FPS process; (ii) photons in the signal and idler
bands produced by the RS process; and (iii) residual pump photons
that leak into the signal and idler bands because of spectral
broadening caused by SPM. Since the Kerr nonlinearity is
relatively weak, for 300\,m of DSF only about 0.1 photons on
average are scattered by a typical 5-ps-duration pump pulse
containing approximately $10^7$ photons. We achieve the required
isolation by sending all the photons emerging from the output port
through a specially made filter F, which is realized in two ways.
In one configuration (DGFAWG), F is composed of a double-grating
filter (DGF), providing an isolation greater than
75\,dB~\cite{Fiorentino02}, followed by array-waveguide-gratings
(AWGs) (Wavesplitter, model WAM-10-40-G), providing isolation
greater than 40\,dB~\cite{Li05c}, wherein the passband of the
DGFAWG is determined by both the passband of the DGF and that of
the channel used in the AWG. The second configuration of F (CWDMF)
is composed of two cascaded WDM filters (JDSU, models
DWS-2F2863P90 and DWS-2F3353P90), each one providing an isolation
greater than 70\,dB in each channel. The two configurations are
schematically shown in Figs.~\ref{setup}(b) and~\ref{setup}(c),
respectively, and their transmission spectra are plotted in
Fig.~\ref{spectra}(a). Each passband of DGFAWG fits a Gaussian
function well with a full-width at half-maximum (FWHM) of 0.4\,nm,
as shown in Fig.~\ref{spectra}(b). In contrast, each passband of
CWDMF is well fitted with a forth-order super-Gaussian function
having a FWHM of 1\,nm, as shown in Fig.~\ref{spectra}(c). The
dynamic range of the spectra shown in Fig.~\ref{spectra} is
limited by the intrinsic noise of the optical spectrum analyzer
(OSA) used to make the measurements. The total isolation to
out-of-band pump photons provided by the filter F in both
configurations is greater than 110\,dB.

The pump is a 5-ps-duration mode-locked pulse train with a
repetition rate of $75.3$\,MHz, obtained by spatially dispersing
the output of an optical parametric oscillator (OPO) (Coherent,
model Mira-OPO) with a diffraction grating~\cite{Li05a}. To
achieve the required power, the pump pulses are then amplified by
an erbium-doped fiber amplifier (EDFA). Photons at the signal and
idler wavelengths---from the OPO that leak through the
spectral-dispersion optics and from the amplified spontaneous
emission in the EDFA---are suppressed by passing the pump through
a 1nm-bandwidth tunable filter (Newport, model TBF-1550-1.0). In
our experiment, the spectrum of pump pulses is well fitted with a
Gaussian function of 0.8\,nm FWHM, as shown in
Fig.~\ref{spectra}(d).

The signal and idler photons are detected with photon counters
consisting of InGaAs/InP avalanche photodiodes (APDs) (Epitaxx,
model EPM 239BA) operated in a gated-Geiger
mode~\cite{Fiorentino02}. The 1-ns-wide gate pulses arrive at a
rate of $588$\,kHz, which is $1/128$ of the repetition rate of the
pump pulses. The quantum efficiency for one detector is $25\%$,
for the other is 20\%. The total detection efficiencies for the
signal and idler photons are about $3.5\%$ and $2.8\%$ ($10\%$ and
$8\%$ with CWDMF), respectively, when the efficiencies of the DSF
($85\%$), DGFAWG ($\simeq 20\%$) [CWDMF ($\simeq 80\%$)], and
other transmission components (about $85\%$) are included. For the
CWDMF, an efficiency of $\simeq 85\%$ should also be included when
PAs are inserted.

\section{Results}

\begin{figure}
\includegraphics[width=3.25in]{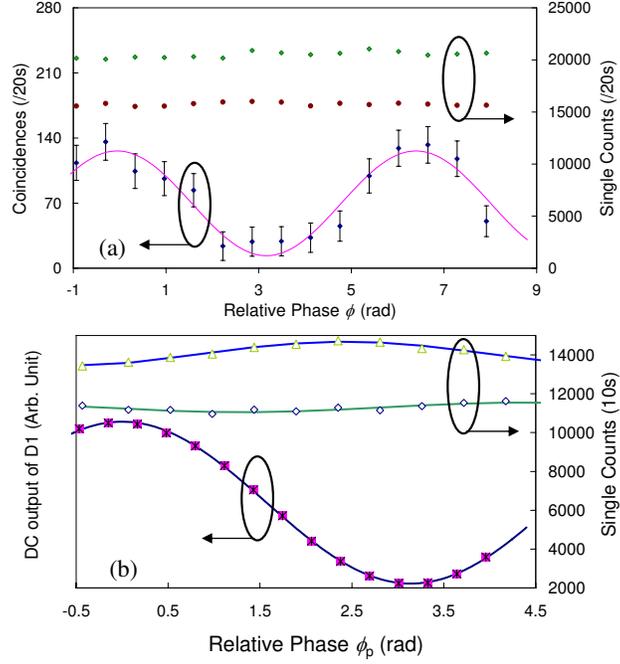}
\caption{\label{interference1} Measurement of polarization
entanglement as the relative phase $\phi _p$ is varied with both
the PAs set at $45^{\circ }$ relative to the verticle. Solid
curves are best fits. (a) Coincidence counts and single counts
detected over 20\,s versus $\phi = 2\phi_p$ when the pump powers
for both $P_H$ and $P_V$ are about 0.3\,mW. (b) Left ordinate:
plot of the output from detector D$_1$ (squares and dark blue
curve) showing one-photon interference in the rejected pump pulses
with twice the fringe spacing as in (a). Right ordinate: single
counts in signal (triangles) and idler (diamonds) bands detected
over 10\,s versus $\phi_p$ when the pump powers for both $P_H$ and
$P_V$ are about 0.4\,mW. The fitting curve for the signal (light
blue) [idler (green)] is phase shifted by about 2 radians behind
(abead) that for the pump.}
\end{figure}

Our first measurement for characterizing the polarization
correlation of the source involves setting both PAs at $45^{\circ
}$ and slowly scanning the relative phase $\phi_p $ between $P_H$
and $P_V$ while observing the photon counts, both singles and
coincidences. Phase scanning is accomplished by adding separate
free-space propagation paths for $P_H$ and $P_V$ with use of a PBS
(P$_5$), two quarter-wave plates (QWP$_2$ and QWP$_3$), and two
mirrors (M$_2$ and M$_3$) [see Fig.~\ref{setup}(d)]. M$_3$ is
mounted on a piezoelectric-transducer-(PZT)-driven translation
stage, which allows us to precisely adjust the free-space path
difference and hence the relative phase $\phi_p $. For this
measurement, the pump powers for both $P_H$ and $P_V$ are about
0.3\,mW and the DGFAWG is used to reject the pump. During the
measurement, the phase $\phi_p $ is simultaneously monitored by
tapping off the rejected pump with a mirror M$_1$ placed after the
grating G$_1$ and measuring it with a low-bandwidth photodetector
D$_1$ placed behind a polarizer P$_2$. This arrangement is
equivalent to monitoring one output port of the polarization
interferometer formed between P$_1$ and P$_2$ [see
Fig.~\ref{setup}(b)]. The fringe contrast of the interfering $P_H$
and $P_V$ pulses is maximized by appropriately adjusting the
position of M$_3$ (by applying an appropriate voltage on the PZT)
and the orientations of HWP$_1$ and QWP$_1$. This ensures that
$P_H$ and $P_V$ arrive at $P_1$ simultaneously, and that the
rejected $P_H$ and $P_V$ are oriented at $\pm 45^{\circ }$
relative to the axis of P$_2$. The results are presented in
Fig.~\ref{interference1}(a), where one sees good interference in
the coincidence counts whereas no interference is observed in the
single counts. In plotting the coincidence counts only the dark
counts of the detectors have been subtracted. By fitting the
plotted coincidence counts with a cosine function, we obtain a TPI
visibility of $71\%$. These results are obtained at a production
rate of 0.1 photons/pulse, which is deduced from the detected
single counts, and among them the photon-pair production rate is
about 0.07 pairs/pulse.

The output of D$_1$ is plotted in Fig.~\ref{interference1}(b),
which when compared with the TPI result shown in
Fig.~\ref{interference1}(a) clearly shows that the relative phase
$\phi$ between the wave packets $| H\rangle_s| H \rangle_i$ and $|
V \rangle_s | V \rangle_i$ is twice the pump phase $\phi_p$, i.e.,
$\phi=2\phi_p $. It is worth noting that when the pump power is
higher, slight variations also begin to appear in the single
counts, with the variation in signal band being more than that in
idler band. This is also shown in Fig.~\ref{interference1}(b),
which was obtained with pump powers for both $P_H$ and $P_V$ to be
about $0.4$\,mW. Fitting the single counts with a cosine function,
we find that the periodicity in the single-count variations is the
same as that of the pump, but the phases are different. For the
signal counts the fringe maximum occurs behind that of the pump by
about 2 radians, while for the idler counts it is ahead of the
pump maximum by the same amount of phase shift. We see a similar
behavior upon further increasing the pump power; only the
magnitude of the variations in the single counts is increased.

We believe that the SPM of the pump is responsible for the
observed variations in the single counts. Although the isolation
provided by the filter F (DGFAWG for the data in
Fig.~\ref{interference1}) is good enough to effectively reject the
pump photons at the central wavelength, the SPM induced spectral
broadening of the pump pulses at higher powers causes pump photons
at the signal and idler passband frequencies to leak through the
filter F. These leakage photons give rise to first-order
interference observed in the single counts.

One way to reduce the influence of the SPM, is to increase the
detuning of the signal band from the pump. However, the detuning
needs to be decreased to suppress the RS in the DSF~\cite{Li04}.
Therefore, we improve the quality of our source by optimizing the
system parameters such that the RS and SPM are suppressed while
the correlation of the photon-pairs produced by the FPS is
maximized, in accordance with our measurements of these effects in
Refs.~\onlinecite{Li04,Chen05}, and~\onlinecite{Li05c}. Thus, we
replace the Gaussian-shaped DGFAWG with a FWHM of 0.4\,nm by the
supper-Gaussian--shaped CWDMF with a FWHM of 1\,nm. Additionally,
because the loss of each WDM used to construct the CWDMF is only
about 0.5\,dB, the efficiency of the source is significantly
improved.

\begin{figure}
\includegraphics[width=3.25in]{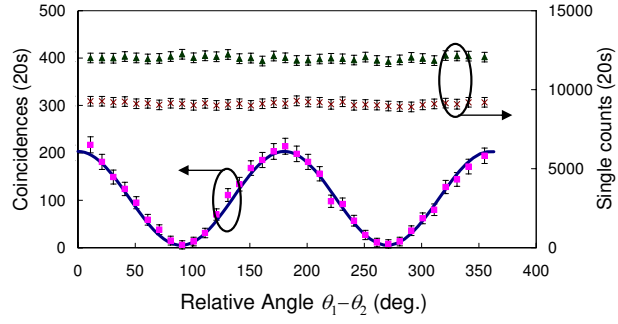}
\caption{\label{interference2} Coincidence counts and single
counts detected over 20\,s as the analyzer angle in the idler
channel is varied while keeping the analyzer angle in signal
channel fixed at 45$^\circ$ relative to the vertical. The solid
curve is a best fit, yielding a TPI visibility of 92\%.}
\end{figure}

With the optimized configuration of the system, we measure the TPI
again with pump powers of $P_H$ and $P_V$ at about $0.05$\,mW
each. Blocking the light reflected from M$_2$ and adjusting
HWP$_4$ to $22.5^{\circ }$ relative to P$_1$ (see
Fig.~\ref{setup}), we obtain a pump which is linearly polarized
with an angle of $45^{\circ }$ relative to P$_1$. In making the
measurement, the QWP$_4$ is either taken out or oriented with its
axis parallel to that of the HWP$_4$. In this way we set $\phi_p
=0$ so that the two-particle quantum state emerging from the
output port is $| \Psi \rangle^{+} =| H\rangle _s| H\rangle _i+|
V\rangle _s| V\rangle _i$. TPI measurements are made by fixing the
PA in the signal channel at $45^{\circ }$ while rotating the PA in
the idler channel. The results are shown in
Fig.~\ref{interference2}, wherein only the dark counts of the
detectors have been subtracted. These results are obtained for a
production rate of about 0.02 photons/pulse (deduced from the
single-count measurements), among which the photon-pair production
rate is about 0.006 pairs/pulse. Clearly, we observe much improved
TPI, while the single counts in both the channels remain constant.
Again by fitting the coincidence counts with a cosine function, a
TPI visibility of $92\%$ is obtained.

For comparison, we increase the pump powers of $P_H$ and $P_V$ to
about 0.15\,mW and make the TPI measurement again. Under this
condition, we obtain a deduced production rate of about 0.13
photons/pulse, among which the photon-pair production rate is
about 0.07 pairs/pulse. Thus, at the same photon-pair production
rate as in Fig.~\ref{interference1}, the RS in this case is
larger. And yet, we observe TPI with a higher visibility of about
78\% than the visibility of 71\% in Fig.~\ref{interference1}. This
increase in visibility is because the correlation of the detected
photons produced via FPS is maximized by using a super-Gaussian
filter such as the CWDMF with a wider and flat passband.

\section{Conclusions}

We have demonstrated a high-quality source of polarization
entangled photon-pairs by using a counter propagating scheme in
which an optical-fiber loop is simultaneously pumped with
orthogonally polarized pulses in clockwise and counter-clockwise
directions. We produce high-quality photon-pairs by carefully
optimizing the system parameters in order to suppress the negative
effects of Raman scattering and self-phase modulation of the pump
pulses, and to maximize the photon-pair correlation through the
four-photon scattering process. The source is integrated with
off-the-shelf telecom components and is, therefore, well suited
for quantum communication applications in the 1550\,nm telecom
band. We have characterized the source by investigating two-photon
interference between the generated signal-idler photon-pairs under
various experimental conditions. At a production rate of about
0.02 photons/pulse in the signal (idler) channel, two-photon
interference with visibility $>90$\% is obtained without
subtracting counts caused by the background photons (only dark
counts of the detectors are subtracted), while no interference is
observed in direct detection of either the signal or the idler
photons. All four Bell states can be created with this source. The
results can be significantly improved by cooling the fiber to
further suppress the Raman scattering. Moreover, the free-space
components of the present setup can be replaced with commercially
available fiber-spliced components, which would make the source
very compact and easily integrable into the existing fiber
network.

\begin{acknowledgments}
We would like to thank Dr. Jay E. Sharping for useful discussions.
This work was supported in part by the DoD Multidisciplinary
University Research Initiative (MURI) Program administered by the
Army Research Office under Grant DAAD19-00-1-0177.
\end{acknowledgments}



\begin{thebibliography}{17}
\expandafter\ifx\csname
natexlab\endcsname\relax\def\natexlab#1{#1}\fi
\expandafter\ifx\csname bibnamefont\endcsname\relax
  \def\bibnamefont#1{#1}\fi
\expandafter\ifx\csname bibfnamefont\endcsname\relax
  \def\bibfnamefont#1{#1}\fi
\expandafter\ifx\csname citenamefont\endcsname\relax
  \def\citenamefont#1{#1}\fi
\expandafter\ifx\csname url\endcsname\relax
  \def\url#1{\texttt{#1}}\fi
\expandafter\ifx\csname
urlprefix\endcsname\relax\def\urlprefix{URL }\fi
\providecommand{\bibinfo}[2]{#2}
\providecommand{\eprint}[2][]{\url{#2}}

\bibitem[{\citenamefont{Bennett and Shor}(1998)}]{Bennett98}
\bibinfo{author}{\bibfnamefont{C.~H.} \bibnamefont{Bennett}} \bibnamefont{and}
  \bibinfo{author}{\bibfnamefont{P.~W.} \bibnamefont{Shor}},
  \bibinfo{journal}{IEEE Trans. Inf. Theory} \textbf{\bibinfo{volume}{44}},
  \bibinfo{pages}{2724} (\bibinfo{year}{1998}).

\bibitem[{\citenamefont{Burnham and Weinberg}(1970)}]{Weinberg70}
\bibinfo{author}{\bibfnamefont{D.~C.} \bibnamefont{Burnham}} \bibnamefont{and}
  \bibinfo{author}{\bibfnamefont{D.~L.} \bibnamefont{Weinberg}},
  \bibinfo{journal}{Phys. Rev. Lett.} \textbf{\bibinfo{volume}{25}},
  \bibinfo{pages}{84} (\bibinfo{year}{1970}).

\bibitem[{\citenamefont{Kwiat et~al.}(1995)\citenamefont{Kwiat, Mattle,
  Weinfurter, Zeilinger, Sergienko, and Shih}}]{Kwait95}
\bibinfo{author}{\bibfnamefont{P.~G.} \bibnamefont{Kwiat}},
  \bibinfo{author}{\bibfnamefont{K.}~\bibnamefont{Mattle}},
  \bibinfo{author}{\bibfnamefont{H.}~\bibnamefont{Weinfurter}},
  \bibinfo{author}{\bibfnamefont{A.}~\bibnamefont{Zeilinger}},
  \bibinfo{author}{\bibfnamefont{A.~V.} \bibnamefont{Sergienko}},
  \bibnamefont{and} \bibinfo{author}{\bibfnamefont{Y.~H.} \bibnamefont{Shih}},
  \bibinfo{journal}{Phys. Rev. Lett.} \textbf{\bibinfo{volume}{75}},
  \bibinfo{pages}{4337} (\bibinfo{year}{1995}).

\bibitem[{\citenamefont{Bovino et~al.}(2003)\citenamefont{Bovino, Varisco,
  Colla, Castagnoli, Giuseppe, and Sergienko}}]{Bovino03}
\bibinfo{author}{\bibfnamefont{F.~A.} \bibnamefont{Bovino}},
  \bibinfo{author}{\bibfnamefont{P.}~\bibnamefont{Varisco}},
  \bibinfo{author}{\bibfnamefont{A.~M.} \bibnamefont{Colla}},
  \bibinfo{author}{\bibfnamefont{G.}~\bibnamefont{Castagnoli}},
  \bibinfo{author}{\bibfnamefont{G.~D.} \bibnamefont{Giuseppe}},
  \bibnamefont{and} \bibinfo{author}{\bibfnamefont{A.~V.}
  \bibnamefont{Sergienko}}, \bibinfo{journal}{Opt. Commun.}
  \textbf{\bibinfo{volume}{227}}, \bibinfo{pages}{343} (\bibinfo{year}{2003}).

\bibitem[{\citenamefont{Castelletto et~al.}(2004)\citenamefont{Castelletto,
  Degiovanni, Migdall, and Ware}}]{Castelletto04}
\bibinfo{author}{\bibfnamefont{S.}~\bibnamefont{Castelletto}},
  \bibinfo{author}{\bibfnamefont{I.~P.} \bibnamefont{Degiovanni}},
  \bibinfo{author}{\bibfnamefont{A.}~\bibnamefont{Migdall}}, \bibnamefont{and}
  \bibinfo{author}{\bibfnamefont{M.}~\bibnamefont{Ware}}, \bibinfo{journal}{New
  J. Phys.} \textbf{\bibinfo{volume}{6}}, \bibinfo{pages}{Art.\ No.\ 87}
  (\bibinfo{year}{2004}).

\bibitem[{\citenamefont{Andrews et~al.}(2004)\citenamefont{Andrews, Pike, and
  Sarkar}}]{Andrews04}
\bibinfo{author}{\bibfnamefont{R.}~\bibnamefont{Andrews}},
  \bibinfo{author}{\bibfnamefont{E.~R.} \bibnamefont{Pike}}, \bibnamefont{and}
  \bibinfo{author}{\bibfnamefont{S.}~\bibnamefont{Sarkar}},
  \bibinfo{journal}{Opt.\ Express} \textbf{\bibinfo{volume}{12}},
  \bibinfo{pages}{3264} (\bibinfo{year}{2004}).

\bibitem[{\citenamefont{Fiorentino et~al.}(2002)\citenamefont{Fiorentino, Voss,
  Sharping, and Kumar}}]{Fiorentino02}
\bibinfo{author}{\bibfnamefont{M.}~\bibnamefont{Fiorentino}},
  \bibinfo{author}{\bibfnamefont{P.~L.} \bibnamefont{Voss}},
  \bibinfo{author}{\bibfnamefont{J.~E.} \bibnamefont{Sharping}},
  \bibnamefont{and} \bibinfo{author}{\bibfnamefont{P.}~\bibnamefont{Kumar}},
  \bibinfo{journal}{IEEE Photon. Technol. Lett.} \textbf{\bibinfo{volume}{14}},
  \bibinfo{pages}{983} (\bibinfo{year}{2002}).

\bibitem[{\citenamefont{Li et~al.}(2005{\natexlab{a}})\citenamefont{Li, Voss,
  Sharping, and Kumar}}]{Li05a}
\bibinfo{author}{\bibfnamefont{X.}~\bibnamefont{Li}},
  \bibinfo{author}{\bibfnamefont{P.~L.} \bibnamefont{Voss}},
  \bibinfo{author}{\bibfnamefont{J.~E.} \bibnamefont{Sharping}},
  \bibnamefont{and} \bibinfo{author}{\bibfnamefont{P.}~\bibnamefont{Kumar}},
  \bibinfo{journal}{Phys. Rev. Lett.} \textbf{\bibinfo{volume}{94}},
  \bibinfo{pages}{053601} (\bibinfo{year}{2005}{\natexlab{a}}).

\bibitem[{\citenamefont{Takesue and Inoue}(2004)}]{Takesue04}
\bibinfo{author}{\bibfnamefont{H.}~\bibnamefont{Takesue}} \bibnamefont{and}
  \bibinfo{author}{\bibfnamefont{K.}~\bibnamefont{Inoue}},
  \bibinfo{journal}{Phys. Rev. A} \textbf{\bibinfo{volume}{70}},
  \bibinfo{pages}{031802} (\bibinfo{year}{2004}).

\bibitem[{\citenamefont{Li et~al.}(2004)\citenamefont{Li, Chen, Voss, Sharping,
  and Kumar}}]{Li04}
\bibinfo{author}{\bibfnamefont{X.}~\bibnamefont{Li}},
  \bibinfo{author}{\bibfnamefont{J.}~\bibnamefont{Chen}},
  \bibinfo{author}{\bibfnamefont{P.~L.} \bibnamefont{Voss}},
  \bibinfo{author}{\bibfnamefont{J.}~\bibnamefont{Sharping}}, \bibnamefont{and}
  \bibinfo{author}{\bibfnamefont{P.}~\bibnamefont{Kumar}},
  \bibinfo{journal}{Opt. Exp.} \textbf{\bibinfo{volume}{12}},
  \bibinfo{pages}{3737} (\bibinfo{year}{2004}).

\bibitem[{\citenamefont{Rarity et~al.}(2005)\citenamefont{Rarity, Fulconis,
  Duligall, Wadsworth, and Russell}}]{Rarity05}
\bibinfo{author}{\bibfnamefont{J.~G.} \bibnamefont{Rarity}},
  \bibinfo{author}{\bibfnamefont{J.}~\bibnamefont{Fulconis}},
  \bibinfo{author}{\bibfnamefont{J.}~\bibnamefont{Duligall}},
  \bibinfo{author}{\bibfnamefont{W.~J.} \bibnamefont{Wadsworth}},
  \bibnamefont{and} \bibinfo{author}{\bibfnamefont{P.~S.~J.}
  \bibnamefont{Russell}}, \bibinfo{journal}{Opt. Express}
  \textbf{\bibinfo{volume}{13}}, \bibinfo{pages}{534} (\bibinfo{year}{2005}).

\bibitem[{\citenamefont{Sharping et~al.}(2004)\citenamefont{Sharping, Chen, Li,
  and Kumar}}]{Sharping04}
\bibinfo{author}{\bibfnamefont{J.~E.} \bibnamefont{Sharping}},
  \bibinfo{author}{\bibfnamefont{J.}~\bibnamefont{Chen}},
  \bibinfo{author}{\bibfnamefont{X.}~\bibnamefont{Li}}, \bibnamefont{and}
  \bibinfo{author}{\bibfnamefont{P.}~\bibnamefont{Kumar}},
  \bibinfo{journal}{Opt. Express} \textbf{\bibinfo{volume}{12}},
  \bibinfo{pages}{3086} (\bibinfo{year}{2004}).

\bibitem[{\citenamefont{Fan et~al.}(2005)\citenamefont{Fan, Dogariu, and
  Wang}}]{Wang05}
\bibinfo{author}{\bibfnamefont{J.}~\bibnamefont{Fan}},
  \bibinfo{author}{\bibfnamefont{A.}~\bibnamefont{Dogariu}}, \bibnamefont{and}
  \bibinfo{author}{\bibfnamefont{L.~J.} \bibnamefont{Wang}},
  \bibinfo{journal}{Opt. Lett} \textbf{\bibinfo{volume}{30}},
  \bibinfo{pages}{1530} (\bibinfo{year}{2005}).

\bibitem[{\citenamefont{Kumar et~al.}(2005)\citenamefont{Kumar, Fiorentino,
  Voss, and Sharping}}]{Kumar05}
\bibinfo{author}{\bibfnamefont{P.}~\bibnamefont{Kumar}},
  \bibinfo{author}{\bibfnamefont{M.}~\bibnamefont{Fiorentino}},
  \bibinfo{author}{\bibfnamefont{P.~L.} \bibnamefont{Voss}}, \bibnamefont{and}
  \bibinfo{author}{\bibfnamefont{J.~E.} \bibnamefont{Sharping}},
  \bibinfo{journal}{U. S. Patent No.\ 6,897,434}  (\bibinfo{year}{2005}).

\bibitem[{\citenamefont{Li et~al.}(2005{\natexlab{b}})\citenamefont{Li, Voss,
  Chen, Sharping, and Kumar}}]{Li05b}
\bibinfo{author}{\bibfnamefont{X.}~\bibnamefont{Li}},
  \bibinfo{author}{\bibfnamefont{P.~L.} \bibnamefont{Voss}},
  \bibinfo{author}{\bibfnamefont{J.}~\bibnamefont{Chen}},
  \bibinfo{author}{\bibfnamefont{J.~E.} \bibnamefont{Sharping}},
  \bibnamefont{and} \bibinfo{author}{\bibfnamefont{P.}~\bibnamefont{Kumar}},
  \bibinfo{journal}{Opt. Lett} \textbf{\bibinfo{volume}{30}},
  \bibinfo{pages}{1201} (\bibinfo{year}{2005}{\natexlab{b}}).

\bibitem[{\citenamefont{Chen et~al.}(2005)\citenamefont{Chen, Li, and
  Kumar}}]{Chen05}
\bibinfo{author}{\bibfnamefont{J.}~\bibnamefont{Chen}},
  \bibinfo{author}{\bibfnamefont{X.}~\bibnamefont{Li}}, \bibnamefont{and}
  \bibinfo{author}{\bibfnamefont{P.}~\bibnamefont{Kumar}},
  \bibinfo{journal}{\pra} \textbf{\bibinfo{volume}{72}},
  \bibinfo{pages}{033801} (\bibinfo{year}{2005}).

\bibitem[{\citenamefont{Li et~al.}(2005{\natexlab{c}})\citenamefont{Li, Voss,
  Chen, Lee, and Kumar}}]{Li05c}
\bibinfo{author}{\bibfnamefont{X.}~\bibnamefont{Li}},
  \bibinfo{author}{\bibfnamefont{P.~L.} \bibnamefont{Voss}},
  \bibinfo{author}{\bibfnamefont{J.}~\bibnamefont{Chen}},
  \bibinfo{author}{\bibfnamefont{K.~F.} \bibnamefont{Lee}}, \bibnamefont{and}
  \bibinfo{author}{\bibfnamefont{P.}~\bibnamefont{Kumar}},
  \bibinfo{journal}{Opt. Express} \textbf{\bibinfo{volume}{13}},
  \bibinfo{pages}{2236} (\bibinfo{year}{2005}{\natexlab{c}}).

\end{thebibliography}

\newpage


\end{document}